# Predicting Gas Well Performance with Decline Curve Analysis: A Case Study on Semutang Gas Field

*Md. Shakil Rahaman[1,*], Ahmed Sakib[2], Ataharuse Samad[3], Md. Ashraful Islam[4]*

[1-4] Department of Petroleum and Mining Engineering, Chittagong University of Engineering & Technology, Chattogram-4349, BANGLADESH

**ABSTRACT**

Decline-curve analysis (DCA) is a widely utilized method for production forecasting and estimating remaining reserves in gas reservoir. Based on the assumptions that past production trend can be mathematically characterized and used to predict future performance. It relies on historical production data and assumes that production methods remain unchanged throughout the analysis. This method is particularly valuable due to its accuracy in forecasting and its broad acceptance within the industry. Wells in the same geographical area and producing from similar geological formations often exhibit similar decline curve parameters. This study applies DCA to forecast the future production performance and estimate the ultimate recovery for the Semutang gas field's well 5 in Bangladesh. Using historical production data, decline curves were generated based on exponential, hyperbolic, and harmonic model equations. The cumulative production estimations were 11,139.34 MMSCF for the exponential model, 11,620.26 MMSCF for the hyperbolic model, and 14,021.92 MMSCF for the harmonic model. In terms of the well's productive life, the estimates were 335.13 days, 1,152 days, and 22,611 days, respectively. Among these models, the hyperbolic decline provided the most realistic forecast, closely aligning with observed production trend. The study highlights the importance of selecting an appropriate decline model for accurate production forecasting and reserve estimation, which is essential for effective reservoir management and resource optimization.

## 1. Introduction

In petroleum engineering, the ultimate recovery, and prediction of the rate at which the production decreases are central tasks undertaken by reservoir engineers. Every well, reservoir, or field on this planet has a finite hydrocarbon volume. Therefore, after a while, an inevitable decline in the rate at which the production decreases occurs. Eventually, this decline would reach an economic limit, and no further production would continue because the productivity of the entire volume is uneconomical. This is the so-called "ultimate recovery." In this case, the cumulative reserves, which are indicated before production commences, will be the ultimate recovery. The cumulative production will always be equal to the ultimate recovery, and the rest will be the remaining reserves. The remaining reserves will continue decreasing whenever something is produced.

A decline curve analysis is known to be one of the most popular methods applied among other methods of reserve estimation. It is more applicable in situations where more historical production information is available. DCA is an empirical modeling technique that enhances the ability to forecast future production levels by extension of previously known production figures and is important when considering the life expectancy of any given well and its production parameters. In DCA, the relationship of production rate with time and production as a whole is an important aspect of this method. Specifically, the established first-order instantaneous rate of production generally exhibits rapid depletion within a very short time scale after which the production slump extends. However, there are some operational factors like changes in production method, equipment, and efficiency, or workover activities that may cause the decline curve to be lost and hence the analysis may need initial smoothing for accuracy.

DCA is popular because of its simplicity and ease to execute, but it assumes future production performance will be similar to the past. This practical, empirical nature does unfortunately mean that these methods may give less accuracy than more cumbersome volumetric calculations [3]. Eventually, when used prudently, DCA provides a powerful tool for analyzing reservoir performance and guiding production strategies and key decisions related to hydrocarbon recovery.

This study will investigate further the methodologies and principles of decline curve analysis (DCA) with Arps's equations [4] applied to reserve estimates and production forecasts in different reservoir conditions. More specifically, the goals of this study are to forecast future production performance and estimate the ultimate recovery.

## 2. Literature review

Decline curve analysis (DCA) is a technique petroleum engineers use to predict future production rates of oil and gas wells. The Arps equations by Arps (1945) lay the very basic foundation of DCA to model production decline at a range of observed performances. The following section discusses some studies in literature that have used Arps' equations for decline analysis.

Three empirical decline model have been suggested; the exponential, the hyperbolic, and the harmonic[4]. Arps, J. J.

---

Specific examples of these models defined by different rates of decline that enable history matching are shown. Arps also emphasized the fact that in using any model for forecasting production it is essential to first understand its mechanisms when considering flow and thus model selection.

Blasingame, D.(2000)[ 5] expanded the original ideas of Arps [2] by investigating the utilization of decline curve analysis in different reservoir behaviors. This says that the model of hyperbolic decline is more realistic and describes production data from reservoirs with different flow types than exponential or harmonic models.

Wiggins, R.A., Baber, M.R.: Evaluating the use of arps' equations for gas well performance prediction in the Barnett shale formation (2003) [6] The elliptical decline type model produced better results in terms of forecast accuracy for producing data from unconventional reservoirs, according to the authors. They also recognized that future recovery estimates were very sensitive to changes in the decline exponent.

Applications of the Arps decline models in older oilfields were studied by Yilmaz, I. (2005) [7]. His results suggested that the hyperbolic model provided a more accurate hypothesis fit to production data complexity, as opposed to traditional exponential methods. The outcomes from this study provided more evidence that the choice of decline model is paramount to robust performance forecasting in mature fields.

Hosseini and Sadeghi (2010) [8]studied Arps decline models along with other forecasting techniques in a carbonate reservoir. These manifestations concluded that Arps' equations are useful for initial estimates, and the inclusion of reservoir-specific parameters is capable of dramatically improving the quality of production forecasts.

Zhang Y, Liu C (2012) [9] The Truth Lies in the Curve: Applying Arps' Equations to Shale Gas Reservoirs. They also stressed the importance of characterizing transient flow behavior in these reservoirs and suggested probable revisions to Arps parameters that would help improve forecast accuracy.

Nikkhah and Nikkhah, 2017 [10] also studied Arps equitation with machine learning techniques. They have proposed that, although Arps models are still sound for retrospective data analysis, a combination of machine learning approaches will provide value to history matching for forecasting in a dynamic reservoir environment.

Combined, these studies demonstrate the adaptability and reliability of Arps' models for decline curve analysis but underscore the importance in using correct model selection procedures depending on reservoir properties. Nevertheless, as the industry progresses, additional work is needed to increase the robustness of these empirical models to different reservoir types and increase functionality by combining with emerging analytical methods which could result in more accurate production forecasts.

### 3. Methodology
3.1 Reservoir description

Semutang is an NNW-SSE elongated anticline located in the Bengal basin's eastern Chittagong fold belt. It partially occupies the districts of Manikchari in Khagrachari and Nazirhat and Fatikchari in Chittagong. About 60 kilometers northeast of Chittagong, a port city, is this location. The Halda and Tinturi synclines form the structure's western and eastern boundaries, respectively. It is approximately 25 kilometers long and 5 kilometers wide. It spans latitudes 22°45'- 22°57' N and longitudes 91°42'- 91°49' E (Fig. 1). Low-lying terrain made up of little hills and hillocks with elevations varying from 100' to 250' covers the surface exposure.

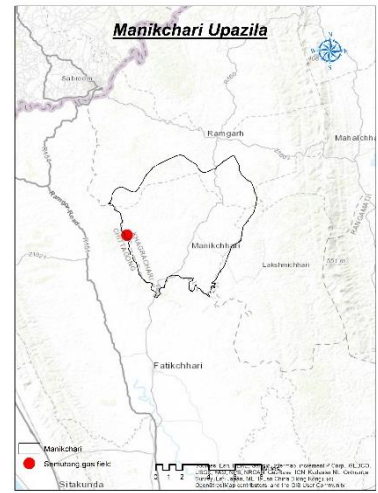

**Fig.1** Digitalized map of the study area

This study describes a step-by-step methodology for forecasting the future performance of a gas well in the reservoir part of the Semutang field using the Arps (1945) [4] model equations. The data used in this study was obtained from the Semutang gas field. This study focuses on projecting gas well performance using decline curve analysis, conducting history matching to evaluate various field development strategies (scenarios), and forecasting future gas output based on the suggested development plan. To perform this analysis, plot the production rate versus time for the exponential, hyperbolic, and harmonic flow models. Figure 2 shows conventional curves for production rate versus time.

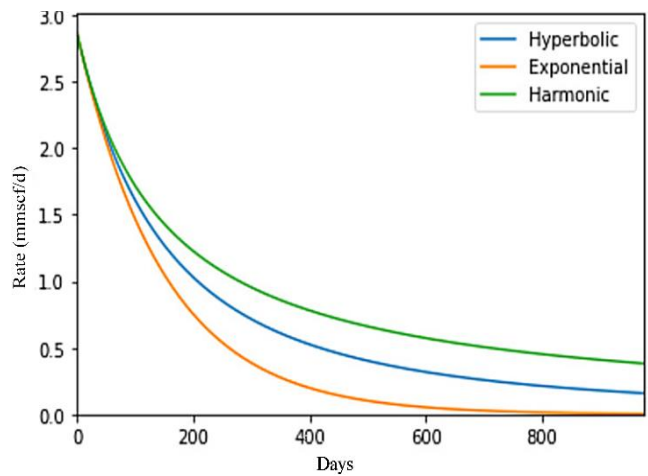

**Fig.2** Arp's decline curves

3.2 Model equation

Production performance datasets from the Semutang field were examined using several equations from the Arps

(1945) [4] model. These equations were applied under the assumptions of pseudo-steady state flow conditions. The decline curve analysis was conducted utilizing these empirical model equations as follows:

3.2.1 Exponential decline analysis

It is a method used in petroleum engineering to model the production rate (refers to the amounts of hydrocarbons (oil, gas, or both) that are produced from a well or reservoir during a certain period. It is usually measured in units like barrels per day for oil or million standard cubic feet per day for gas) decline of oil or gas wells over time. In this approach, the production rate decreases at a constant percentage rate per unit of time. The decline rate $D_i$ (refers to the fractional rate of decrease in production over time. It determines how quickly the production rate decreases as a function of time and is a critical parameter in forecasting future production and estimating reserve) remains constant concerning the production rate, q when factor b (describes the rate at which the decline rate declines with time. Simply put, it determines how "steep" or "gradual" the production curves decrease) is zero. In this case, the production rate q at any given time can be expressed as:

$$q = q_i\, e^{-D_i t} \tag{1}$$

Rate-cumulative production,

$$Q = \frac{q_i - q}{D_i} \tag{2}$$

Excepted time,

$$\Delta t = \frac{1}{D} \ln\left(\frac{q_f}{q_{ab}}\right) \tag{3}$$

3.2.2 Harmonic decline analysis

Here, the production rate q decreases at a progressively slower rate, meaning the decline rate $D_i$ continuously decreases as time progresses. The decline rate changes linearly concerning the production rate when the factor b = 1.
In this case, the production rate q at any given time can be expressed as:

$$q = q_i (1 + D_i t)^{-1} \tag{4}$$

Rate-cumulative production,

$$Q = \frac{q_i}{D_i} \ln\left(\frac{q_i}{q}\right) \tag{5}$$

Excepted time,

$$\Delta t = \frac{q_i}{D_i}\left(\frac{1}{q_{ab}} - \frac{1}{q_f}\right) \tag{6}$$

3.3.3 Hyperbolic decline analysis

The rate of decline is not constant but decreases over time. It is a more flexible approach than exponential or harmonic decline models, allowing for a varying decline rate that better reflects the production behavior of many reservoirs. The decline rate Di varies geometrically with the production rate, q when the factor is 0 < b < 1. In this case, the production rate q at any given time can be expressed as:

$$q = q_i(1 + bD_i t)^{\frac{-1}{b}} \tag{7}$$

Rate-cumulative production,

$$Q = \frac{q_i}{(1-b)D_i}\left[1 - \left(\frac{q}{q_i}\right)^{1-b}\right] \tag{8}$$

Excepted time,

$$\Delta t = \frac{\left(\frac{q_i}{q}\right)^b - 1}{b \times D_i} \tag{9}$$

Where,
$q_i$ = initial gas production rate, mmscf/d
q = gas production rate, mmscf/d
$D_i$ = nominal decline rate, /day
t = decline period, day
$N_p$ = cumulative gas production, mmscf

## 4. Results and discussion
4.1 Exponential decline curve analysis

The production data curves in Fig. 3 for Well 5 are essential for Exponential Decline Curve Analysis (DCA). The Cartesian rate-time plot shows the production rate decline over time, while the semi-log rate-time plot depicts exponential declines as straight line. The Cartesian rate-cumulative production plot correlates production rate with cumulative output, aiding in the interpretation of production trend and exponential decline pattern.

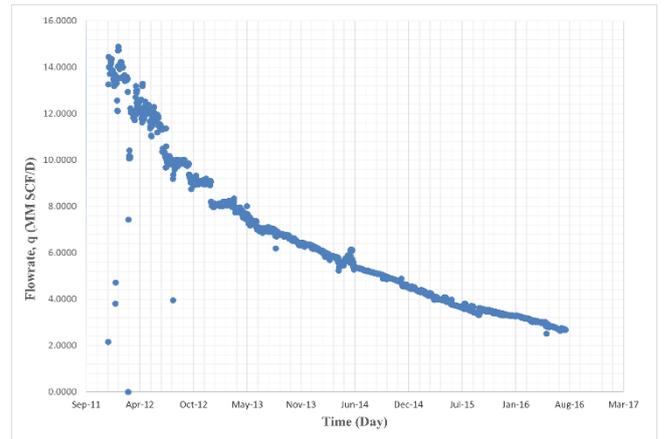

(a)

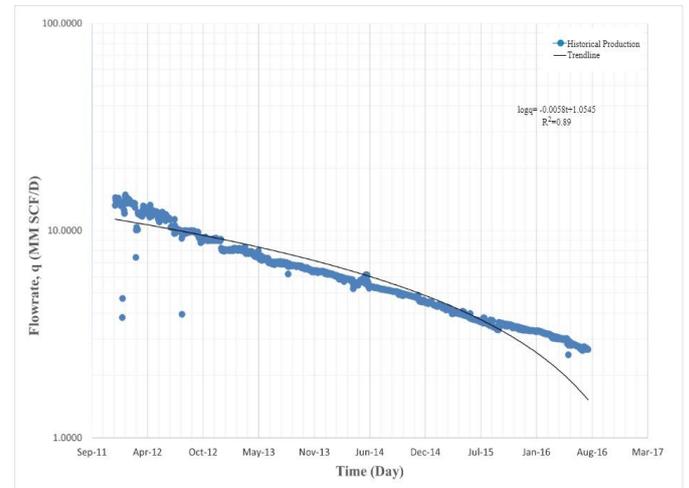

(b)

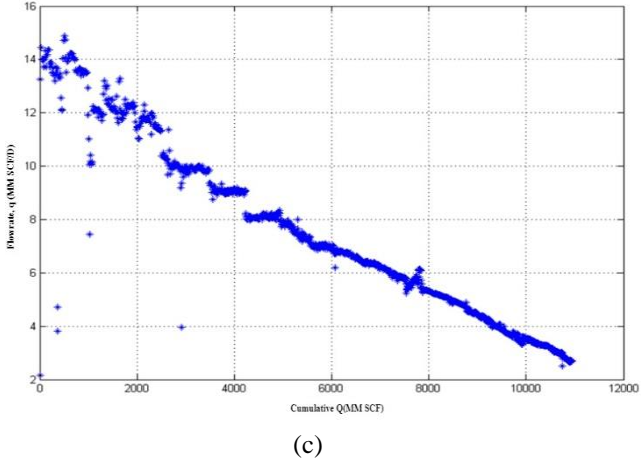

(c)

**Fig.** (a) Production rate vs time in the Cartesian plot, (b) semilog rate vs time plot, (c) rate vs cumulative production plot

Based on the linear regression results of Fig.3 (b) it is evident that the decline rate is almost constant which does not change concerning time. Finally, it is obtained

$$\log q = 1.0545 - 0.0058\, t \tag{10}$$

This can be expressed as,

$$q = 11.339\, e^{-0.0134t} \tag{11}$$

Based on the production data of well 5, the current production rate is 2.6755 mmscf/d, and the flow rate at the time of abandonment is $q_{ab}$ = 0.03 mmscf/d or 30 mscf/d. The cumulative production $Q_f$ from the equation (2) is estimated as follows

$$Q_f = \frac{q_f - q_{ab}}{D}$$
$$= \frac{2.6755 - 0.03}{0.0134}$$
$$= 197.4 \text{ mmscf}$$

The ultimate recoverable volume is estimated as follows and the cumulative production, is 10,941.9205 mmscf, from Fig.3 (c).

EUR = $N_p + Q_f$
  = (10941.9205 + 197.4) mmscf
  = 11139.34587 mmscf

Finally, the production time reaching the abandonment from equation (3) is

$$\Delta t = \frac{1}{D} \ln \left(\frac{q_f}{q_{ab}}\right)$$
  = (1/0.0134) ln (2.6755/0.03)
  = 335.13 day

The analysis shows that Well 5 is expected to produce an ultimate recoverable volume (EUR) of approximately 11,139.35 mmscf. With a current production rate of 2.6755 mmscf/d and an abandonment rate of 0.03 mmscf/d, the well is projected to continue producing for around 335 days.

4.2 Harmonic decline curve analysis

The production rate of a well was plotted against time using a Cartesian coordinate system to determine the initial decline rate for harmonic decline curve analysis in Fig.4.

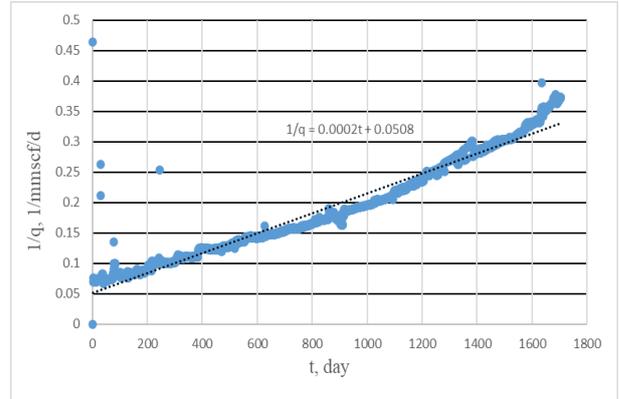

**Fig.4** Cartesian rate countdown–time plot

Based on the linear regression results of Fig.4, we have

1/q = 0.0002t + 0.0508 (12)

By rearranging this equation into,

$$q = \frac{q_i}{(1+D_i t)} = 19.69/\,(1+0.0039t) \tag{13}$$

Comparing this equation with equation (4), getting Di is 0.0039 per day. Now from equation (5), the cumulative production $Q_f$ is estimated as follows:

$$Q_f = \frac{q_i}{D_i} \ln \frac{q_f}{q_{ab}}$$

  = (2.6755/0.0039) × ln (2.6755/0.03) mmscf

  = 3080 mmscf

The ultimate recoverable volume is

EUR = $N_p + Q_f$

  = (10941.9205 + 3080) mmscf

  = 14021.9205 mmscf

Based on equation (6), the production time reaching the abandonment is

$$\Delta t = \frac{q_i}{D_i}\left(\frac{1}{q_{ab}} - \frac{1}{q_f}\right)$$

  = (2.6755/0.0039) × {(1/0.03) − (1/2.6755)}

  = 2611 days

The analysis indicates that Well 5 is anticipated to yield an ultimate recoverable volume (EUR) of roughly 14,021.92 mmscf. With a current production rate of 2.6755 mmscf/d and an abandonment rate of 0.03 mmscf/d, the well is expected to remain productive for approximately 22,611 days.

4.3 Hyperbolic decline analysis

Hyperbolic decline occurs when the decline rate decreases over time. The degree of this decrease is constant and defined by the "b" factor. Fig.4 shows a clear bend and the amount of bend in this curve depends on the b factor: a higher b factor results in a greater bend. Additionally, the steepness of the curve is determined by the initial decline rate of production.

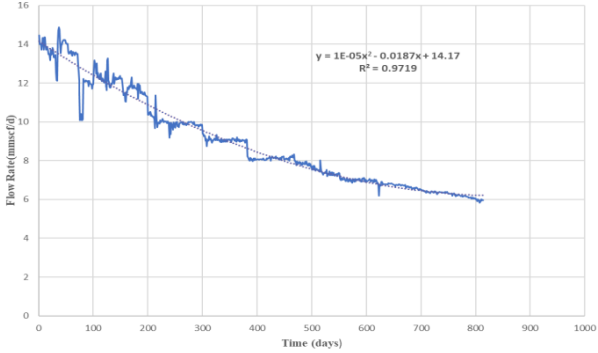

**Fig.5** Hyperbolic Decline with Decreasing Rate Over Time.

The straight line in Fig.6 illustrates a consistent decline rate that represents the b factor. The value for this decline rate is calculated to be $2\times10^{-5}$ based on the graph.

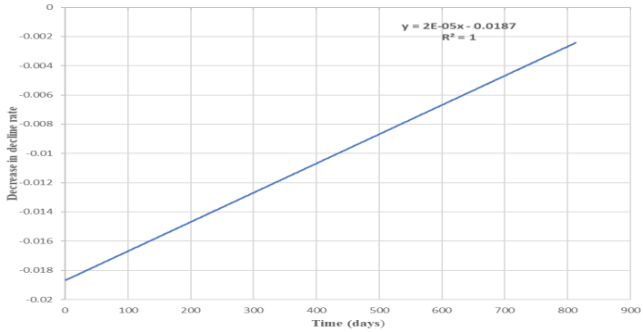

**Fig.6** Calculating factor 'b'

Similarly, from equation (8) using the equation for rate cumulative production is calculated as,

$Q = \frac{q_i}{(1-b)D_i}[1-(\frac{q}{q_i})^{1-b}]$

$= \frac{2.6755}{(1-2\times10^{-5})\times0.0039}[1-(\frac{0.03}{2.6755})^{1-2\times10^{-5}}]$

$= 678.34$ mmscf

The ultimate recoverable volume is,

EUR= $N_p + Q_f$

= (10941.9205 + 678.34) mmscf

=11620.26mmscf

Based on equation (9), the production time reaching the abandonment is

$\Delta t = \frac{(\frac{q_i}{q})^b - 1}{b \times D_i}$

$= \frac{(\frac{2.6755}{0.03})^{2\times10^{-5}} - 1}{2\times10^{-5}\times0.0039}$

=1152 days

The analysis indicates that Well 5 is projected to yield an ultimate recoverable volume (EUR) of about 11,620.26 mmscf. Given its current production rate of 2.6755 mmscf/d and an abandonment re of 0.03 mmscf/d, the well is expected to remain productive for approximately 1,152 days.

Finally, the results for the predicted outcomes of exponential decline, harmonic decline, and hyperbolic decline are presented in Table 4

**Table 4** EUR and Δt for different decline curve methods

| Methods | Exponential | Harmonic | Hyperbolic |
|---|---|---|---|
| EUR(mmscf) | 11139.34587 | 14021.9205 | 11620.26 |
| Δt (day) | 335.13 | 22611 | 1152 |

The following Fig.7 illustrates the production forecast derived from the analysis of historical production data through Python analysis with an $R^2$ value of 0.90 and RMSE value of 0.60. The results indicate that the well is projected to remain productive for approximately 1,200 days, ultimately reaching an abandonment flow rate of approximately 0.3 million standard cubic feet per day (mmscf/day).

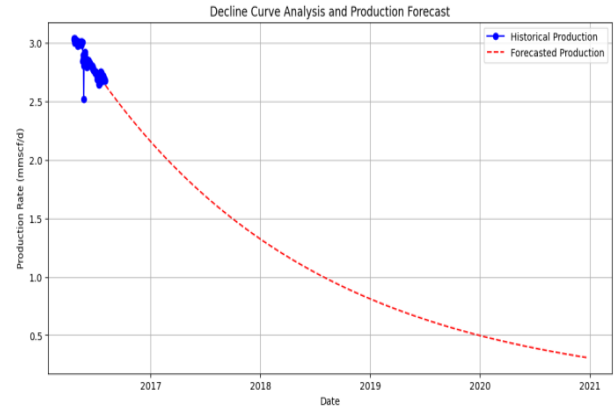

**Fig.7** Production Forecast for Well 5 Based on Historical Data Analysis

The production history of Well-5 in the Semutang gas field shows that as of August 6th, 2017, the well had a production rate of 2.7655 mmscf/d. A later production record indicates the well reached its abandonment flow rate on January 18th, 2021, meaning the well remained productive for approximately 1,257 days. Hyperbolic Decline Curve Analysis (DCA) forecasted a productive lifespan of about 1,152 days for the well to reach abandonment flow from 2.7655 mmscf/d, closely aligning with the actual production history of 1,257 days. Additionally, simulation results (Fig. 6) predicted that the well would continue producing until 2021, confirming the production history. Moreover, from Fig. 4 and Fig. 5, the calculated 'b factor' value is $2 \times 10^{-5}$, satisfying the condition 0 < b < 1 according to equation 7 of the hyperbolic DCA method, which also supports the method's accuracy. Thus, among the three decline curve analysis methods, hyperbolic DCA demonstrated the highest forecasting accuracy, with a prediction accuracy of 92%, surpassing the other two methods.

## 5. Limitations

- The decline curve analysis is effective during a reservoir's boundary-dominated flow phase. It is much less reliable at the beginning of the well production life, called the transient flow phase.
- Over the years, DCA assumes the parameters that influence the production, such as reservoir pressure fluid characteristics, to be constant. Such variables are, however, prone to changes in practice, which may affect the forecasts' degree of accuracy.

## 5. Conclusion

In conclusion, production data for decline curve analysis must be obtained with care, as the extent of recoverable hydrocarbons are inherently uncertain. Using multiple estimation methods, helps to mitigate this uncertainty by allowing for a comparison of results. Predicting future production is challenging, as recovery will only be known when a well is abandoned. Even after initial estimates, operating companies may refine these forecasts to strengthen confidence over time.

- This study employed exponential, hyperbolic, and harmonic decline models to estimate cumulative production and productive life.
- The hyperbolic model provided the most realistic forecast, with cumulative production estimates of 11,620.26 MMSCF and a productive life of 1,152 days, closely aligning with observed data.
- The production prediction was obtained by examining historical production data using Python, which had an $R^2$ value of 0.90 and an RMSE value of 0.60.

Eventually, while accurate data is essential, the expertise of professionals conducting the analysis is equally important. Companies should ensure high standards in both data acquisition and analysis to achieve reliable production forecasts.

## 7. References

[1] T. Marhaendrajana and T. A. Blasingame, "Decline curve analysis using type curves — evaluation of well performance behavior in a multiwell reservoir system," *SPE Annual Technical Conference and Exhibition*, Sep. 2001, doi: 10.2118/71517-ms.

[2] H. Pratikno, J. A. Rushing, and T. A. Blasingame, "Decline curve analysis using type curves — fractured wells," *SPE Annual Technical Conference and Exhibition*, Oct. 2003, doi: 10.2118/84287-ms.

[3] D. Ilk, J. A. Rushing, A. D. Perego, and T. A. Blasingame, "Exponential vs. Hyperbolic Decline in Tight Gas Sands — Understanding the Origin and Implications for Reserve Estimates Using Arps' Decline Curves," *All Days*, Sep. 2008, doi: 10.2118/116731-ms.

[4] J. J. Arps, *Analysis of decline curves*, vol. 160. *Transactions of the AIME*, 1945.

[5] D. Blasingame, *Decline curve analysis for unconventional reservoirs*. SPE Annual Technical Conference and Exhibition, 2000.

[6] R. a. Wiggins and M. R. Baber, *Application of Arp's equations for gas well performance prediction in the Barnett Shale*. SPE Annual Technical Conference and Exhibition, 2003.

[7] Yilmaz, *Evaluation of Arp's decline models in mature oil fields*. SPE Annual Technical Conference and Exhibition, 2005.

[8] S. H. Hosseini and M. Sadeghi, *Comparative study of Arp's decline models and alternative forecasting methods in carbonate reservoirs*. SPE Annual Technical Conference and Exhibition, 2010.

[9] Y. Zhang and C. Liu, *Using Arp's equations for decline curve analysis in shale gas reservoirs*, vol. 5. Journal of Natural Gas Science and Engineering, 2012.

[10] F. Nikkhah and R. Nikkhah, *Integrating Arp's equations with machine learning for reservoir forecasting*, vol. 159. Journal of Petroleum Science and Engineering, 2017.

## NOMENCLATURE

DCA: Decline curve analysis
EUR: Ultimate recoverable volume
NNW: North-Northwest
SSE: South-Southeast
$\Delta t$ : Production time reaching the abandonment, day
$q_{ab}$ : flowrate at abandonment, mmscf/d